\documentclass[a4paper,11pt]{article}
\usepackage{aaskaiid}

\usepackage{orcidlink}
\usepackage[shortlabels]{enumitem}
\usepackage[utf8]{inputenc}
\usepackage{graphicx}
\usepackage{subcaption}
\usepackage{ulem}
\usepackage{acronym}

\acrodef{3g}[3G]{third-generation}
\acrodef{bao}[BAO]{baryon acoustic oscillations}
\acrodef{bbh}[BBH]{binary black hole}
\acrodef{bns}[BNS]{binary neutron star}
\acrodef{cbc}[CBC]{compact binary coalescence}
\acrodef{CDM}[$\Lambda$CDM]{$\Lambda$ Cold Dark Matter}
\acrodef{CE}[CE]{Cosmic Explorer}
\acrodef{CMB}[CMB]{Cosmic Microwave Background}
\acrodef{DEMNUni}[DEMNUni]{Dark Energy and Massive Neutrino Universe}
\acrodef{desi}[DESI]{Dark Energy Spectroscopic Instrument}
\acrodef{dm}[DM]{dark matter}
\acrodef{em}[EM]{electromagnetic}
\acrodef{ET}[ET]{Einstein Telescope}
\acrodef{FoF}[FoF]{Friends-of-Friends}
\acrodef{grb}[GRB]{gamma-ray burst}
\acrodef{gw}[GW]{gravitational wave}
\acrodef{kn}[KN]{kilonova}
\acrodef{hi}[HI]{neutral hydrogen}
\acrodef{im}[IM]{intensity mapping}
\acrodef{kn}[KN]{kilonova}
\acrodef{los}[LoS]{line-of-sight}
\acrodef{lss}[LSS]{large-scale structure}
\acrodef{lvk}[LVK]{LIGO-Virgo-KAGRA}
\acrodef{ns}[NS]{neutron star}
\acrodef{nsbh}[NSBH]{neutron star -  black hole}
\acrodef{sne}[SNe]{supernovae}
\acrodef{skao}[SKAO]{SKA Observatory}
\acrodef{snr}[SNR]{signal-to-noise ratio}

\newcommand{\dd}{{\rm d}}

\newcommand{\hi}{\textrm{H\textsc{i}}}

\title{Cosmology from Synergies Between SKAO Surveys and Gravitational Wave Observations}
\ShortTitle{Cosmology-GW Synergies}

\ShortName{Baker et al.} 

\author[1]{Tessa Baker\orcidlink{0000-0001-5470-7616}}
\author[2,3]{Alberto Colombo\orcidlink{0000-0002-7439-4773}}
\author[1,4]{Steven Cunnington\orcidlink{0000-0001-6594-107X}}
\author[5]{Ulyana Dupletsa\orcidlink{0000-0003-2766-247X}}
\author[6,7,8]{José Fonseca\orcidlink{0000-0003-0549-1614}}
\author[9]{Ian Harrison\orcidlink{0000-0002-4437-0770}}
\author[10]{Konstantin Leyde\orcidlink{0000-0001-7661-2810}}
\author[2]{Simone Mastrogiovanni\orcidlink{0000-0003-1606-4183}}
\author[11,12,13]{Riccardo Murgia\orcidlink{0000-0002-2224-7704}}
\author[14]{Dounia Nanadoumgar-Lacroze\orcidlink{0009-0009-7255-8111}}
\author[15,16]{Tommaso Ronconi\orcidlink{0000-0002-3515-6801}}
\author[16,17]{Giulio Scelfo\orcidlink{0009-0002-9028-6147}}
\author[18,19]{Matteo Schulz\orcidlink{0009-0005-8184-0232}}
\author[20,8]{Marta Spinelli\orcidlink{0000-0003-0148-3254}}

\affiliation[1]{Institute of Cosmology \& Gravitation, University of Portsmouth, Dennis Sciama Building, Portsmouth, PO1 3FX, UK}
\affiliation[2]{INFN, Sezione di Roma, 1-00185 Roma, Italy}
\affiliation[3]{INAF -- Osservatorio Astronomico di Brera, via Emilio Bianchi 46, I-23807 Merate (LC), Italy}
\affiliation[4]{Jodrell Bank Centre for Astrophysics, Department of Physics \& Astronomy, out The University of Manchester, Manchester M13 9PL, UK}
\affiliation[5]{Marietta Blau Institute - Austrian Academy of Sciences, 1010 Vienna, Austria}
\affiliation[6]{Departamento de F\'isica e Astronomia, Faculdade de Ci\^{e}ncias, Universidade do Porto, CAUP, Porto, Rua do Campo Alegre 687, Porto, 4169-007, Portugal}
\affiliation[7]{Instituto de Astrof\'isica e Ci\^encias do Espa\c{c}o, Universidade do Porto CAUP, 4150-762 Porto, Portugal}
\affiliation[8]{Department of Physics \& Astronomy, University of the Western Cape, Cape Town 7535, South Africa}
\affiliation[9]{School of Physics and Astronomy, Cardiff University, CF24 3AA, UK}
\affiliation[10]{Center for Computational Astrophysics, Flatiron Institute, 162 5th Ave, New York, NY 10010}
\affiliation[11]{Dipartimento di Fisica, Universit\`a degli Studi di Cagliari, Cittadella Universitaria, 09042 Monserrato (CA), Italy}
\affiliation[12]{INFN, Sezione di Cagliari, Cittadella Universitaria, 09042 Monserrato (CA), Italy}
\affiliation[13]{INAF - Osservatorio Astronomico di Cagliari, Via della Scienza 5, 09047 Selargius (CA), Italy}
\affiliation[14]{Institut de Física d’Altes Energies (IFAE), The Barcelona Institute of Science and Technology, Campus UAB, E-08193 Bellaterra, Barcelona, Spain} 
\affiliation[15]{INAF - Institute of Radioastronomy (IRA), Via Gobetti 101, 40129 Bologna, Italy}
\affiliation[16]{SISSA, Via Bonomea 265, 34136 Trieste, Italy}
\affiliation[17]{INFN, Sezione di Trieste, Via Valerio 2, I-34127 Trieste, Italy}
\affiliation[18]{Gran Sasso Science Institute (GSSI), Viale F. Crispi 7, L'Aquila (AQ), I-67100, Italy}
\affiliation[19]{INFN - Laboratori Nazionali del Gran Sasso (LNGS), L'Aquila (AQ), I-67100, Italy}
\affiliation[20]{Observatoire de la Côte d’Azur, Laboratoire Lagrange, Bd de l’Observatoire, 06304 Nice, France}

\abstract{Both radio surveys and gravitational wave observations have the potential to probe unprecedented volumes in the Universe with systematics independent of existing cosmological probes. This gives them the potential, especially in combination, to solve some of the outstanding issues in cosmology. Here we show how the use of radio observations as electromagnetic signals tracing large scale structures can be used to provide information on gravitational wave \textit{standard sirens}. By providing a way to link luminosity distance to redshift, this combination can constrain the expansion history of the Universe, and provide information on the Hubble constant ($H_0$). We describe the utility of SKA-Mid neutral hydrogen (\hi) intensity maps, \hi\ galaxy redshift surveys, and continuum galaxy surveys compared to and in conjunction with large contemporaneous optical surveys. Such radio tracers provide a unique, large volume and high redshift information for gravitational wave source, capable of probing effectively the expansion history of the Universe $H(z)$.}

\begin{document}
\maketitle

\section{Introduction}
The discovery of the Universe’s expansion by Hubble~\citep{Hubble1929} stands as a landmark achievement in astrophysics. Since then, determining the Hubble constant has remained a central objective in cosmology. As the measurements became increasingly precise, a persistent and unresolved discrepancy, known as the Hubble tension, emerged~\citep{DiValentino:2021izs}.

This tension arises from the difference between two measurement approaches: indirect estimates derived from early-Universe \ac{CMB} observations and direct measurements based on local-Universe type Ia \ac{sne} data. These manifest a $\sim 5\sigma$ tension: we get $H_0 = 67.49 \pm 0.53\,\text{km s}^{-1}\text{Mpc}^{-1}$ from the \ac{CMB} measurements~\citep{Planck2018} and $H_0 = 73.04 \pm 1.04\,\text{km s}^{-1}\text{Mpc}^{-1}$ from \ac{sne}~\citep{Riess:2021jrx}.

The first direct detection of \acp{gw}~\citep{LIGOScientific:2016vbw} has opened the way to a new and independent probe of the expansion, which may someday shed light on the tension.

\Acp{gw} from \acp{cbc}, \acp{bbh}, \acp{bns}, and \acp{nsbh} are the only astrophysical sources we measured so far for which we are able to access directly the luminosity distance, without having to rely on any calibration or distance ladder, representing one of the main shortcomings of the \textit{standard candle} cosmology. The idea of GW cosmology dates back to the seminal paper by~\cite{Schutz}. \Acp{gw}, in fact, can be used as \textit{standard sirens}~\citep{Holz:2005df,DelPozzo:2011vcw}. Being cosmic rulers, \acp{gw} provide us with a self-calibrated measurement of the luminosity distance. Access to the redshift information, on the other hand, is not straightforward. 
Since in GW cosmology constraining the expansion rate of the Universe requires an independent measurement of both distance and redshift, alongside the luminosity distance derived from gravitational measurements, there is the need to develop methods to associate the redshift information.

The easiest and most informative approach requires the identification of an \ac{em} counterpart. When at least one of the two components of a \ac{cbc} merger is a \ac{ns}, there is the possibility that part of the ejecta material fuels an \ac{em} counterpart, either in the form of a \ac{kn} or a \ac{grb}, or both. In such an event, the unique identification of the host galaxy, when the observation of the \ac{em} counterpart is feasible, allows easy access to the redshift association to the \ac{gw} event. In such a case, we speak of \textit{bright sirens}. So far, there has been only one bright siren measurement, coming from the merger of a \ac{bns}, for which we observed both the \ac{kn} and the \ac{grb}, GW170817~\citep{LIGOScientific:2017adf}. The $H_0$ constraint from GW170817, measuring $H_0 = 70^{+12}_{-8}\ \mathrm{km\,s^{-1}\,Mpc^{-1}}$ (at $68\%$\,C.L. highest density interval), is in agreement with both the measurements from Planck 2018 data survey~\citep{Planck2018} and Type Ia \ac{sne} constraints~\citep{Riess:2021jrx}, although with large uncertainties. The $H_0$ measurement from GW170817 still prevent us from deeper insights into the Hubble, tension, but this type of events would allow, in principle, sub-percent precision in constraining $H_0$ ~\citep{Chen:2017rfc}. 

Interestingly, it is possible to measure the Hubble parameter even without an \ac{em} counterpart: the increasing number of such \ac{cbc} observations, most of them being \acp{bbh}, allows us to exploit \textit{dark siren} methods. A particular version of this method, often referred to as the \textit{catalogue} method, combines the information on luminosity distance and sky localisation available from the detected GW waveform with the redshift information available from a galaxy catalogue \citep{Holz:2005df,DelPozzo:2011vcw,Gray:2021sew,Gair:2022zsa}, with the combined redshift information across the catalogue used as a \textit{redshift prior} for \ac{cbc} events. For each GW source a probabilistic association is made with the galaxies as potential host locations, corresponding to a given redshift and a given $H_0$. Though for a typical sky localisation regions for an individual GW event there will be many possible associated galaxy redshifts (and hence many possible $H_0$ values), the likelihoods for many individual GW events may be multiplied together, with the probability mass at the correct $H_0$ value being coherently reinforced and other noise-like associations being diminished. To date, this method has mostly involved photometric optical and near-Infrared galaxy surveys due to their relative depth and sky coverage, ensuring the greatest probability that the correct host galaxy is included.

Additionally, the \textit{spectral sirens} method also uses dark sirens events, but instead of applying a redshift prior formed from the positions of detected potential host galaxies, it instead uses the expected joint mass-redshift distribution of \ac{cbc} events determined from population modelling \citep[e.g.][]{Ezquiaga:2022zkx}. Similarly, though this may lead to highly biased results for small numbers of GW events, the correct $H_0$ value may be increasingly isolated as orders of magnitude more sources are included. This prior information may also be combined with the prior information on \ac{cbc} redshift from the catalogue method, with this often being referred to as a \textit{line of sight redshift prior}.

This method comes with several complications, including the correct relative normalisation of host probability for each potential galaxy position. This involves not only the redshift probability distribution for the galaxy ($P_{\rm gal}(z)$) but the probability that a particular galaxy is a host of a particular \ac{cbc} ($P_{\rm host}$). These two probabilities may depend on many of the same underlying variables (e.g. galaxy mass, metallicity) requiring a careful treatment to avoid biases on the cosmological inference \citep{Mastrogiovanni:2021wsd}. 
Another complication arises in the inclusion of completeness of the galaxy catalogue as a function of redshift as one needs to correctly account for the probability that the \ac{cbc} has occurred in a galaxy which is not part of the galaxy catalogue \citep[see e.g.][for a description]{Gray_2020}.

The \ac{lvk} Collaboration has reported constraints on the Hubble constant using either features in the mass spectrum alone, \textit{spectral sirens}, or in combination with the information coming from statistical association with galaxy catalogs~\citep{LIGOScientific:2019zcs, LIGOScientific:2021aug, LIGOScientific:2025jau}. 

Looking ahead, the future \ac{gw} observatories, such as the \ac{ET}~\citep{Punturo:2010zz, Maggiore:2019uih, Branchesi:2023mws, ET:2025xjr} and \ac{CE}~\citep{Reitze:2019iox, Evans:2021gyd, Gupta:2023lga}, will open a new observational era. These \ac{3g} detectors will provide an order-of-magnitude increase in sensitivity compared to the current \ac{lvk} network, enabling the detection of sources out to very high redshifts and producing an order of $\sim 10^5$ CBC events per year. Such a wealth of data will make population-level studies and statistical approaches to cosmology with \acp{gw} increasingly powerful. They will also begin to detect \ac{cbc} at increasing redshifts beyond the reach of optical galaxy catalogues.

In this context, a new synergy between \acp{gw} and other \ac{em} tracers of the \ac{lss} is emerging. Of particular interest here is the ability of neutral hydrogen (\hi) \ac{im} \citep[e.g.][]{Battye:2004re,Santos_2017,SKA:2018ckk} to provide us with redshift information to be combined the luminosity distance measurement coming from GWs. We will describe two uses. First, the three dimensional map of the matter in the Universe offered by \hi\ \ac{im} can be used, similarly to what is currently done with galaxy catalogues, to provide a \ac{los} redshift prior. Secondly, the cosmological parameters are inferred from the cross-power spectrum of maps of the spatial distributions of \hi\ and \ac{gw} sources.

\begin{figure}
    \centering
    \includegraphics[width=0.75\linewidth]{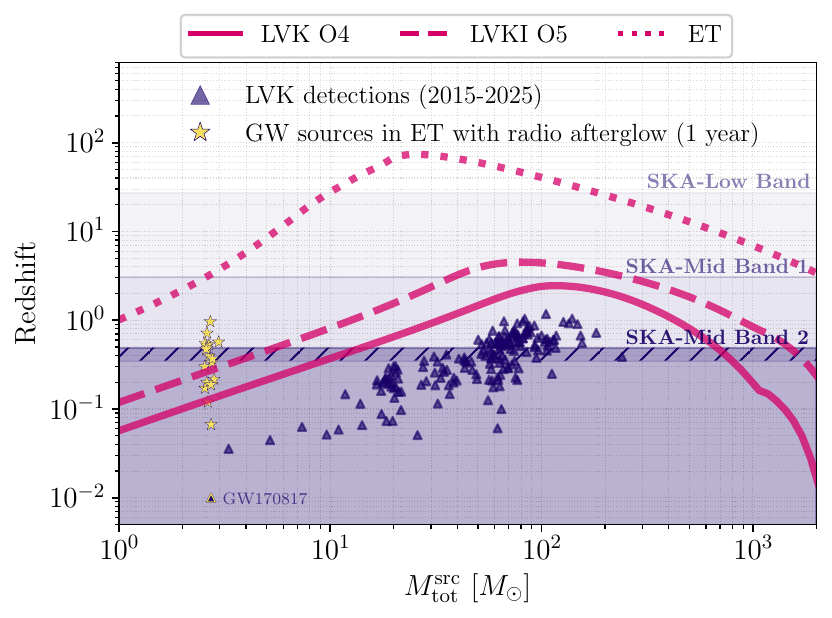}
    \caption{Horizon plots (\textit{magenta} lines) for current and future \ac{gw} observations compared to the \ac{skao} reach in redshift. The horizon represents the maximum redshift for \ac{cbc} equal-mass and optimally oriented \ac{gw} sources observed with a \ac{snr} threshold of 8. On the \textit{x-axis} is the total source-frame mass of the binary, while on \textit{y-axis} is the redshift of the merger. The different line styles refer to different detectors' networks. The \textit{solid} line refers to the \ac{lvk} network during the ongoing fourth observing run, using an optimistic-case simulated sensitivity curve. The \textit{dashed} line refers to the future \ac{lvk} upgrade comprising the LIGO India interferometer. The \textit{dotted} one represents the leap achievable with a \ac{3g} detector like the \ac{ET}. The sensitivity curves for the \ac{gw} detectors are publicly available \href{https://dcc.ligo.org/LIGO-T2200043/public}{here} for the \ac{lvk} detectors and \href{https://apps.et-gw.eu/tds/?r=18213}{here} for \ac{ET} (see footnote for details). The current catalog of \ac{gw} sources, comprising observations from 2015 till 2025, GWTC-4.0~\citep{LIGOScientific:2025slb} is represented with \textit{purple triangles}, while a simulation of one year of \ac{ET}-detectable \ac{bns} sources with an associated afterglow in the radio band, detectable with \ac{skao}, is marked with \textit{yellow triangles}. We highlighted GW170817, our only bright siren event. The \ac{skao} redshift ranges are highlighted horizontally in \textit{purple}: there are the two bands of SKA-Mid, band 2 ($950-1760$\,MHz) and band 1 ($350-1050$\,MHz), and the high-redshift SKA-Low band ($50-350$\,MHz). The intersection region of the two SKA-Mid bands is hatched.}
    \label{fig:horizon_plot}
\end{figure}

\ac{skao} will be capable of surveying large volumes of the Universe across much of the sky with redshift ideally spanning the range $0.35 < z \lesssim 3$ with SKA-Mid but also $3 \lesssim z \lesssim 20$ with SKA-Low, potentially providing redshift prior information for dark sirens analyses in regions unavailable to the traditional optical catalogues. The redshift ranges for the different SKAO telescope bands are shown as the purple regions of Fig.~\ref{fig:horizon_plot}, along with the observable horizons of future GW observatories. This represents a unique opportunity, not only for increasing precision on $H_0$, but also to probe the expansion history $H(z)$ at higher redshift, allowing  to test models of modified gravity~\cite{Ishak:2024jhs} and dynamical dark energy~\cite{DESI:2024kob,DESI:2025fii}. Compared to galaxy surveys, which typically become highly incomplete at redshifts above $\sim 1$, \ac{im} from SKA-Mid could survey the entire souther sky up to $z \sim 3$ \citep{SKA:2018ckk}, making \ac{im} an ideal complement which can provide redshift information for GW sources which would otherwise lie outside the informative part of the redshift prior. This use of \hi\ information in dark sirens analysis is discussed extensively in Sec.~\ref{subsec:hi_los}.

In Fig.~\ref{fig:horizon_plot}, we are showing a summary of the synergy between \ac{3g} \ac{gw} observatories and observable with \ac{skao}. We use the horizon as a metric for \ac{gw} capabilities. This figure of merit shows, as a function of the total source-frame binary mass, the maximum redshift that such a system would be observed with a \ac{snr} threshold of eight under ideal conditions, i.e., equal-mass optimally oriented binaries. We show different network capabilities, in \textit{magenta}, using different line styles. The \textit{solid} line refers to the \ac{lvk} network during the ongoing fourth observing run, using an optimistic-case simulated sensitivity curve. The \textit{dashed} line refers to the future \ac{lvk} upgrade comprising the LIGO India interferometer. The \textit{dotted} one represents the leap achievable with a \ac{3g} detector like the \ac{ET}\footnote{The sensitivity curves for the \ac{gw} detectors are publicly available \href{https://dcc.ligo.org/LIGO-T2200043/public}{here} for the \ac{lvk} detectors and \href{https://apps.et-gw.eu/tds/?r=18213}{here} for \ac{ET}. In particular, we use \textit{aligo$\_$O4high.txt} for the LIGO O4 sensitivity and \textit{AplusDesign.txt} for its O5 upgrade. We use \textit{avirgo$\_$O4high$\_$NEW.txt} for the O4 stage of Virgo, and \textit{avirgo$\_$O5high$\_$ NEW.txt} for O5. Finally, for KAGRA O4 sensitivity, we employ the \textit{kagra$\_$10Mpc.txt} curve,  and \textit{kagra$\_$128Mpc.txt} for the O5 stage. For \ac{ET}, we employ the full cryogenic sensitivity for a 10\,km-arm triangle shape as studied in~\cite{Branchesi:2023mws}}. The current catalog of \ac{gw} sources, comprising observations from 2015 till 2025, GWTC-4.0~\citep{LIGOScientific:2025slb} is represented with \textit{purple triangles}.

Finally, as well as dark siren line of sight redshift priors, radio afterglows from \ac{cbc} mergers can also be used to provide individual host identifications, giving $z$ information for estimating $H_0$. Expectations for these afterglows are also shown by the yellow stars in Fig.~\ref{fig:horizon_plot} and will be discussed in detail in Sec.~\ref{subsec:bright_sirens}. The wide redshift range covered with \ac{skao} well matches the potential of future \ac{gw} observatories, not only on the sheer gravitational detection side, but also for the potential radio emissions associated with some categories of \ac{gw} sources, such as mergers involving \acp{ns}.

This chapter is organized as follows. In Sec.~\ref{sec:standard_sirens}, we introduce the general framework of the \ac{los} cosmological approach, discussing both \textit{bright} (Sec.~\ref{subsec:bright_sirens}) and \textit{dark} (Sec.~\ref{subsec:dark_sirens}) sirens. The application of the \hi\ \ac{los} method is then outlined in Sec.~\ref{subsec:hi_los}. In Sec.~\ref{sec:radio_tracers}, we describe additional radio tracers relevant for the \ac{los} approach, including continuum (Sec.~\ref{subsec:continuum_gal}) and \hi\ galaxies (Sec.~\ref{subsec:hi_gal}). The cross-correlation analysis is presented in Sec.~\ref{sec:cross_corr}, and our main conclusions are summarized in Sec.~\ref{sec:conclusions}.

\section{GW sirens with line-of-sight redshift priors}\label{sec:standard_sirens}
\subsection{Current Approaches}
\subsubsection{Bright Sirens}\label{subsec:bright_sirens}

Mergers involving at least one neutron star, such as \ac{bns} and \ac{nsbh} systems, can give rise to a variety of \ac{em} counterparts, including \ac{kn}e and short \acp{grb} with their associated prompt and afterglow emission. When such counterparts are detected, they enable the secure identification of the host galaxy and its redshift, providing the redshift information required to pair with the luminosity distance inferred directly from the \ac{gw} signal. These events, commonly referred to as \textit{bright sirens}, thus offer a completely self-calibrated distance--redshift relation that is independent of any external distance ladder.

As mentioned in the introduction, so far only one bright siren has been detected, GW170817, together with its \ac{kn} and short \ac{grb} counterparts~\citep{LIGOScientific:2017ync,LIGOScientific:2017adf}. The firm identification of its host galaxy, NGC~4993, provided a direct estimate of the Hubble constant. Since then, one additional \ac{bns} event has been detected but with insufficient sky localisation for meaningful \ac{em} follow-up, and none of the \ac{nsbh} systems observed so far have shown detectable \ac{em} emission---likely due to limited mass ejection suppressing bright \ac{kn} or \ac{grb} signatures.

The landscape is expected to change dramatically with the advent of \ac{3g} \ac{gw} observatories, which will increase the detection rate of \ac{bns} and \ac{nsbh} mergers by orders of magnitude. In parallel, the next generation of \ac{em} facilities will enhance our ability to identify their counterparts. Together, these advances will transform bright sirens from rare occurrences into a powerful and statistically rich probe of cosmic expansion, enabling precise cosmological measurements and the possibility of reaching sub-percent accuracy on $H_0$ with sufficiently large samples~\citep{Mancarella:2024qle,Cozzumbo:2024vxw}.

Radio observations will play a central role in this endeavour. As discussed in \citep{GW-GRB-KN-chapter}, radio afterglows from short \ac{grb} can persist far longer than other \ac{em} bands, making radio follow-up a robust route for detecting and characterising \ac{gw}-triggered transients. Radio observations provide unique insights into jet structure and energetics, constrain the viewing angle, and may reveal an afterglow even when a \ac{kn} is too faint to be detected.

In this context, the capabilities of the \ac{skao} are particularly significant. With its high sensitivity, large field of view, and broad frequency coverage, the \ac{skao} will be ideally suited to discovering and monitoring radio afterglows from \ac{bns} and \ac{nsbh} systems. Crucially, even when optical facilities fail to detect a counterpart, \ac{skao} observations can still enable host-galaxy identification and improve constraints on source geometry, thereby refining the luminosity-distance measurement and strengthening the cosmological inference.

We perform here a cosmological-inference study to assess the potential of the bright siren method when combining \ac{3g} \ac{gw} detectors and \ac{skao}. Specifically, we simulate a population of BNS mergers as in \citet{GW-GRB-KN-chapter}, detected in \acp{gw} by \ac{ET} and accompanied by radio afterglows observable with \ac{skao}. Radio counterparts are generated at the central frequency of \ac{skao} band 5a (6.55 GHz), and an event is considered detected if the afterglow peak flux exceeds 8.5 $\mu$Jy. We find 18 detectable GW$+$radio afterglow events in one year of observations. This simulated population of one year of \ac{cbc} sources with an associated afterglow in the radio band and detectable on the gravitational side with \ac{ET}, is marked with \textit{yellow triangles} in Fig.~\ref{fig:horizon_plot}.

In this analysis, we do not explicitly incorporate afterglow-derived constraints on the viewing angle into the GW parameter estimation. Instead, radio observations are used to identify the subset of events with a detectable counterpart, for which a redshift measurement can be obtained and included in the bright-siren analysis.
It is important to note that this approach is a simplification: we do not include sky-localization uncertainties or the impact of astrophysical contaminants that would inevitably affect a realistic \ac{em} follow-up campaign. Moreover, the results depend sensitively on the assumed local \ac{bns} merger rate and \ac{grb} afterglow model, so the predicted detection rate can vary significantly. Despite these caveats,  these forecasts are important to underlying the possible role of \ac{skao} for bright sirens.

\begin{figure}[t!]
    \centering
    
    \begin{subfigure}[t]{0.48\textwidth}
        \centering
        \includegraphics[width=\linewidth]{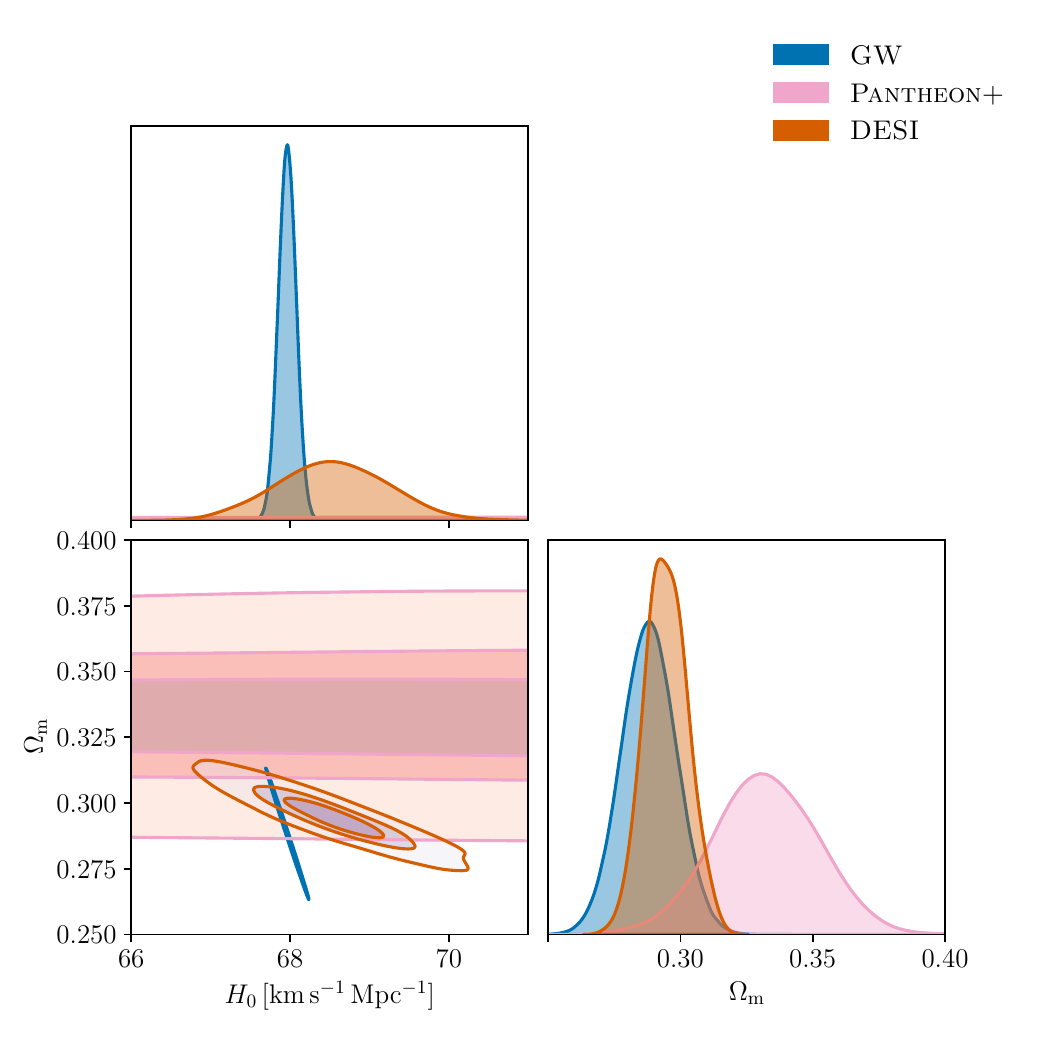}
        \label{fig:1_bright_sir}
    \end{subfigure}
    \hfill
    \begin{subfigure}[t]{0.48\textwidth}
        \centering
        \includegraphics[width=\linewidth]{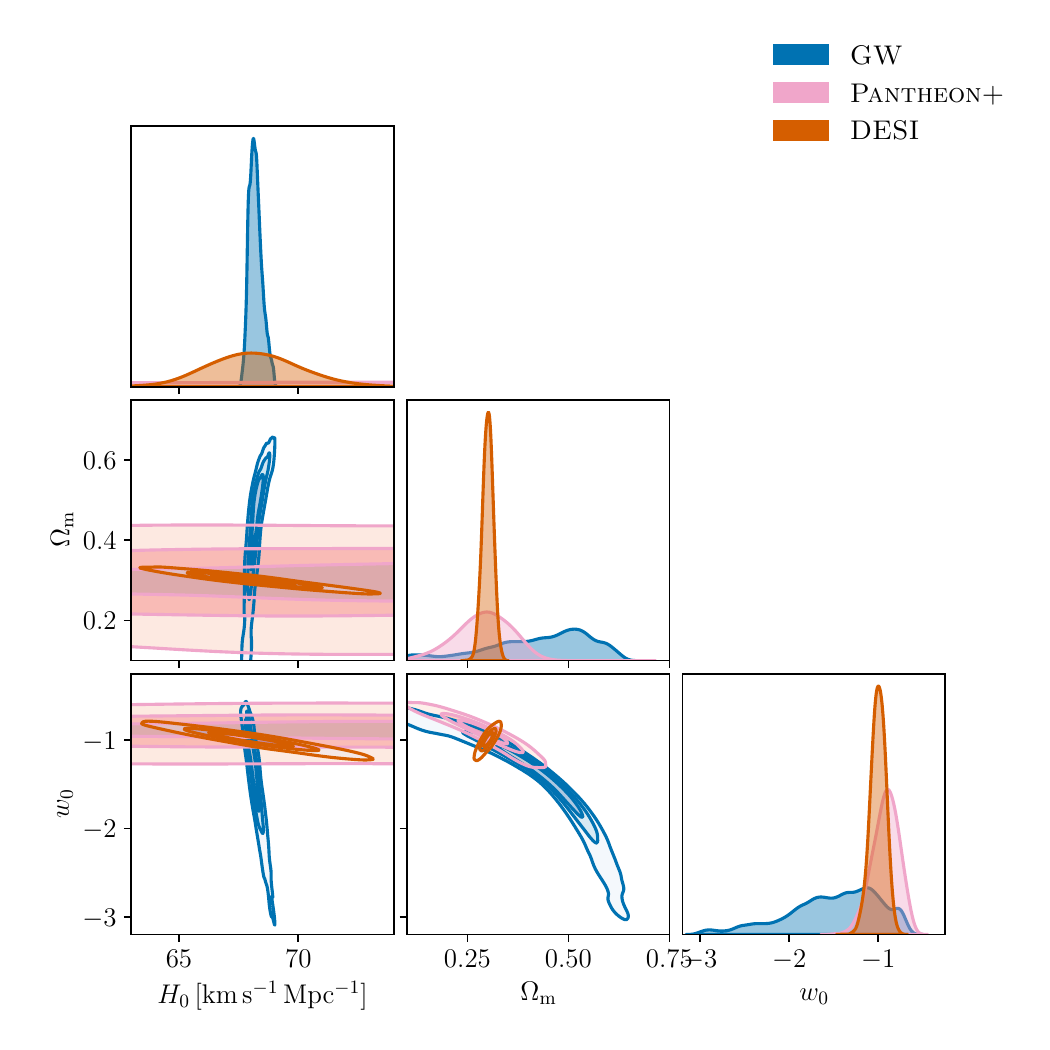}
        \label{fig:2_bright_sir}
    \end{subfigure}

    \vspace{0.2cm} 
    
    \begin{subfigure}[t]{0.6\textwidth}
        \centering
        \includegraphics[width=\linewidth]{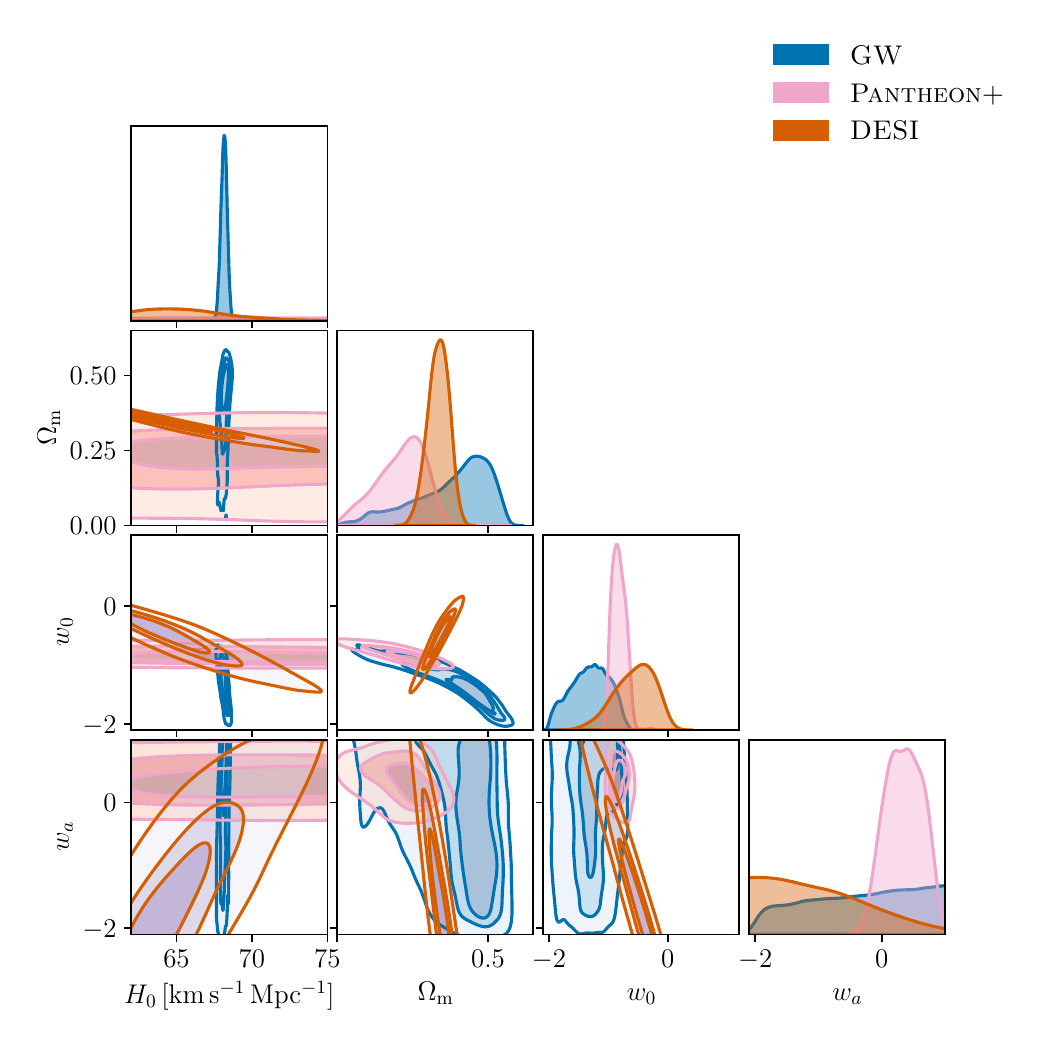}
        \label{fig:3_bright_sir}
    \end{subfigure}

    \caption{Corner plots of the cosmological constraints obtained from \ac{gw} bright sirens (blue), Pantheon\,+ \ac{sne} (pink), and \ac{desi} BAO (orange) for the $\Lambda$CDM model (top left), the $w$CDM model (top right), and the $w_{0}w_{a}$CDM model (bottom). For each model we show the one–dimensional marginal distributions and the two–dimensional posterior contours for the relevant cosmological parameters $H_{0}$, $\Omega_{\mathrm{m}}$, and, when applicable, $w_0$ and $w_{a}$.}
    \label{fig:corner_bright_sirens}
\end{figure}

In Fig.~\ref{fig:corner_bright_sirens}, we report the cosmological constraints obtained from our analysis for three models considered: $\Lambda$CDM (top left), $w$CDM (top right), and $w_{0}w_{a}$CDM (bottom). For each cosmology, we show the marginal posterior distributions and the joint contours for the relevant parameters: $H_{0}$, $\Omega_{\mathrm{m}}$, and, where applicable, $w_{0}$ and $w_{a}$. Three datasets are used in this analysis to illustrate their complementarity: \ac{gw} bright sirens from \ac{skao} follow-up of \ac{ET} events (blue), the uncalibrated type Ia \ac{sne} from Pantheon+ (pink) \cite{Brout:2022vxf}, and \ac{desi} \ac{bao} \cite{DESI:2024mwx} (orange). 
As we can observe from the figure, about 20 bright sirens detected in the \ac{skao}/\ac{ET} era will be able to constrain $H_0$ and the dark matter energy density $\Omega_m$ to a precision significantly better than, or equal to, the one of current measurements from \ac{sne} and \ac{bao}. Instead, in terms of the dark energy Equation of state parameters, we might expect \ac{gw} bright sirens to provide a constraint on $w_0$ and $w_a$, but not as precise as the one currently available from \ac{sne} \cite{Brout:2022vxf}. 
Another interesting aspect of bright sirens is that they do not share any systematic with \ac{sne} or \ac{bao}, as such, these two datasets can be combined to break some of the degeneracies present among cosmological parameters ($H_0$ and $\Omega_m$, for example) and provide a per cent-level precision of some of the cosmological parameters.

\subsubsection{Dark Sirens}\label{subsec:dark_sirens}

Dark siren methods comprise both spectral and catalogue approaches. Spectral sirens rely on the information coming from the astrophysical model of the mass spectrum of \acp{cbc} in the source frame. The mass observed at the detector, in fact, is redshifted, and a model of the mass distribution in the source frame allows us to pinpoint the redshift of a given event:
\begin{equation}
    \label{eq:source_det_mass}
    (1+z(d_L, H(z))) = \frac{m}{m_{\rm src}},
\end{equation}
where $m$ is the detector-frame measured mass, and $m_{\rm src}$ is the mass in the source frame. The presence of certain features in the mass spectrum, such as peaks and breaks, is crucial for inferring the redshift. Peaks indicate the most probable regions in the mass distribution, while breaks signify changes in the distribution. The reliability of this method increases as the amount of data grows.

The statistical framework for dark siren-like analysis is the hierarchical Bayesian inference~\citep{Mandel:2018mve, Vitale:2020aaz}. The inference process operates on two levels. The first concerns the parameter estimation of individual \ac{gw} events, while the second, known as population inference, aims to describe the collective set of observed \ac{gw} events within a coherent framework governed by the mass spectrum, event rate, and cosmological parameters.

The detection of \acp{gw} is an inhomogeneous Poisson process, influenced by selection effects that stem from the finite sensitivity of the detectors. Given $N_{\rm obs}$ \ac{gw} events, collectively specified as $\{x\}$, each of which is described by its own set of parameters $\theta$, the hierarchical likelihood that describes the probability of observing the specific dataset in a given observation time $T_{\rm obs}$, and given the set of population-level parameters $\Lambda$, is~\citep{Mandel:2018mve, Vitale:2020aaz, Mastrogiovanni:2021wsd, Mastrogiovanni:2023emh, Mastrogiovanni:2023zbw}:
\begin{equation}
    \label{eq:hierarchical_lkh}
    \begin{aligned}
    \mathcal{L}(\{x\}|\Lambda) &\propto \exp{\left[-N_{\rm exp}(\Lambda) \right]}\prod_{i=1}^{N_{\rm obs}}\int \mathcal{L}_{\rm GW}\left(x_i|\theta, \Lambda \right)\frac{\dd N_{\rm CBC}}{\dd t \dd\theta}(\Lambda)\dd\theta \dd t\\
    &\propto \exp{\left[-N_{\rm exp}(\Lambda) \right]}\prod_{i=1}^{N_{\rm obs}} T_{\rm obs} \int \mathcal{L}_{\rm GW}\left(x_i|\theta, \Lambda \right)\frac{\dd N_{\rm CBC}}{\dd t \dd \theta}(\Lambda)\dd\theta,
    \end{aligned}
\end{equation}
where, passing from the first to the second line, we have integrated over the observing time, $t$, measured in the detector frame. There are two terms:

\begin{enumerate}
   \item The individual \ac{gw} event likelihood, $\mathcal{L}_{\rm GW}\left(x_i|\theta, \Lambda \right)$, which accounts for how precisely we can measure the binary parameters from \ac{gw} observations;
    
   \item  The differential number of \ac{gw} sources $\frac{\dd N_{\rm CBC}}{\dd t \dd \theta}(\Lambda)$ modelling the probability of the single event parameters $\theta$ given the population-level $\Lambda$. The set of population-level $\Lambda$ comprises parameters modeling the rate $\Lambda_{\rm rate}$, the source intrinsic parameters, as e.g. the masses, $\Lambda_{\rm pop}$, as well as cosmological parameters like $H_0$ and $\Omega_m$, referred to as $\Lambda_{c}$.
    
\item Selection effects enter into the evaluation of the expected number of events $N_{\rm exp}$:    
    \begin{equation}
        N_{\rm exp} (\Lambda) = T_{\rm obs} \int p_{\rm det} (\theta) \frac{\dd N_{\rm CBC}}{\dd t\dd\theta}(\Lambda) \dd \theta,
    \end{equation}
   where $p_{\rm det}$ is the detection probability:
    \begin{equation}
   \label{eq:pdet}
       p_{\rm det}(\theta) = \int_{x\in \text{detectable}} \mathcal{L}_{\rm GW}(x_i|\theta) \dd x,
    \end{equation}
   where the integration is carried out over all the individual \ac{gw} events likelihood in the set $\{x\}$. Evaluating Eq.\eqref{eq:pdet} is not straightforward. The customary approach is to use Monte Carlo simulations of a great number of injected \ac{gw} events, which are simulated based on the chosen population-parameters $\Lambda$ and evaluating the fraction of detected ones. We are using the \ac{snr} in the chosen detector network configuration as the metric to evaluate whether an event is detected or not.
\end{enumerate}

The latter is the core term in hierarchical Bayesian inference. As we will specify in subsequent sections, is the one that can incorporate information of the \ac{los} matter distribution, coming from tracer of the large-scale structure as optical galaxies or \hi.

\begin{figure}
    \centering
    \includegraphics[width=0.75\linewidth]{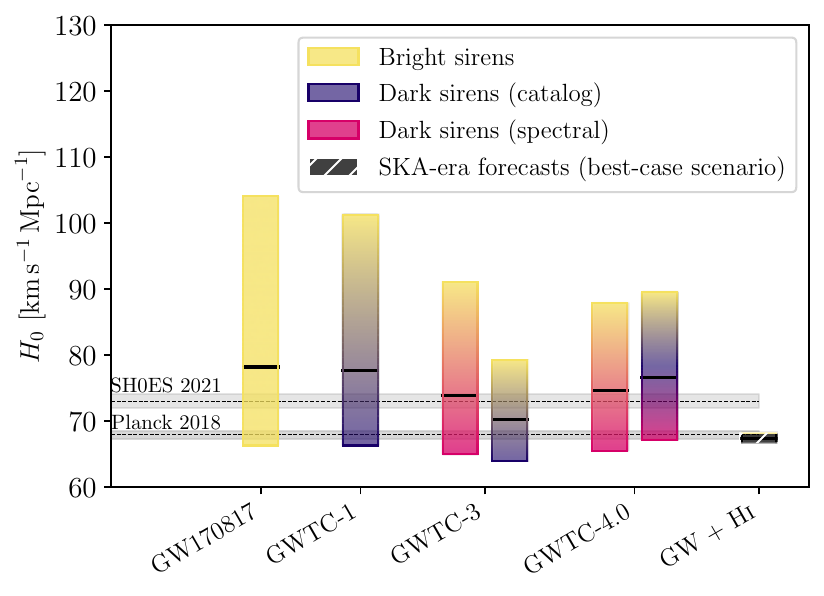}
    \caption{Summary plot of \ac{lvk} $H_0$ measurements~\citep{LIGOScientific:2017adf, LIGOScientific:2019zcs, LIGOScientific:2021aug, LIGOScientific:2025jau} in chronological order. The reported values show the median and the corresponding $68\%$ symmetric credible interval. We summarize the methods employed for the measurements with different colors: \textit{yellow} for bright sirens and either \textit{magenta} (spectral) or \textit{purple} (catalog) for the dark sirens. When more than one method is employed the colors are blended accordingly. We compare these measurements to results from Planck 18~\citep{Planck2018Overview} and SH0ES~\citep{Riess:2021jrx}, for which we show the median value and the $1\sigma$ region. On the gravitational side, note that if the spectral siren method is not included, it means that the population model has been fixed. We compare the evolution of \ac{gw} measurements to the \ac{skao} era forecasts, in the most optimistic scenario. We report the Hubble constant constraint using perfectly measured \ac{gw} as explained in detail in the subsequent Sec.~\ref{subsec:hi_los}.  This figure is adapted from Fig. 11 of~\cite{LIGOScientific:2025jau}, to which we refer for more details.}
    \label{fig:lvk_gw_H0_measurements}
\end{figure}

In Fig.~\ref{fig:lvk_gw_H0_measurements}, we present a summary of GW cosmology measurements as from the \ac{lvk} publications~\citep{LIGOScientific:2017adf, LIGOScientific:2019zcs, LIGOScientific:2021aug, LIGOScientific:2025jau}. The reported values show the median and the corresponding $68\%$ symmetric credible interval. We summarize the methods employed for the measurements with different colors: \textit{yellow} for bright sirens and either \textit{magenta} (spectral) or \textit{purple} (catalog) for the dark sirens. When more than one method is employed the colors are blended accordingly. We compare these measurements to results from Planck 18~\citep{Planck2018Overview} and SH0ES~\citep{Riess:2021jrx}, for which we show the median value and the $1\sigma$ region. On the gravitational side, note that if the spectral siren method is not included, it means that the population model has been fixed.

\subsection{Neutral Hydrogen 21cm Intensity Maps as a Redshift Prior}\label{subsec:hi_los}

An excellent source of complementary information for gravitational-wave cosmology will come from wide-area maps of the cosmic large-scale structure. The SKAO will deliver these through its 21cm intensity mapping surveys, which will trace the three-dimensional distribution of neutral hydrogen (\hi) across cosmic time. Intensity mapping exploits the fact that, on large scales, the aggregate 21cm emission from unresolved galaxies traces the underlying matter density field \cite{Bharadwaj:2000av,Battye:2004re,Chang:2007xk,Wyithe:2007rq}. Instead of detecting individual sources, the telescope measures fluctuations in the collective brightness temperature of the redshifted 21cm line as a function of position and frequency. Each frequency channel directly maps to a cosmological redshift, yielding a tomographic view of the large-scale structure over enormous cosmic volumes.

For this purpose, SKA-Mid will operate in auto-correlation, or \textit{single-dish}, mode \citep{Santos_2017}, where the total power detected by each antenna is used to construct sky maps of the integrated \hi\ emission. This observing mode is essential for accessing the largest angular scales, which are not captured by interferometric visibilities due to their limited baseline lengths. The measurements will use Band 1 of SKA-Mid (350-1050 MHz), corresponding to a redshift range of approximately $0.3\,{<}\,z\,{<}\,3$ encompassing much of the post-reionisation Universe where large numbers of binary black hole mergers detectable by the SKAO-era GW network are expected to occur.

Although the angular resolution of the single-dish observations is modest (${\sim}1^{\circ}$ at the centre of Band 1), the intensity maps provide exquisite radial precision through their spectroscopic frequency resolution. This radial information directly constrains redshift and hence distance, allowing detailed three-dimensional localisation when combined with GW measurements \citep{Oguri:2016dgk,Mukherjee:2022afz}. The combination of wide-field angular coverage with high-fidelity redshift information will enable joint analyses that measure the cosmic distance–redshift relation and expansion rate, $H(z)$, using dark sirens calibrated against the large-scale structure traced by \hi.

The MeerKLASS single-dish survey with the SKA precursor MeerKAT has already started delivering the first results in the L-band (with a redshift range similar to Band 2) \citep{Wang:2020lkn,Cunnington:2022uzo,2025MNRAS.537.3632M} and is paving the way for a larger and deeper survey with the SKA-Mid. A review of the MeerKLASS observational strategy and results can be found in the chapter \citet{Cunnington01.2026.SKA}, while the interested reader can find a summary of the SKA pathfinder results in the chapter \citet{Elahi01.2026.SKA}.

An SKAO \hi\ intensity mapping campaign could ultimately survey up to $20{,}000\,\text{deg}^2$ of sky, making it the largest three-dimensional cosmological map ever constructed \citep{SKA:2018ckk}. In chapter \citet{Spinelli01.2026.SKA}, the challenges linked to foreground cleaning and the most common systematics of an 21cm intensity mapping survey are discussed. \citet{Wolz01.2026.SKA} instead discuss its constraining power for cosmology.

Covering a broad redshift range with precise radial resolution, this survey will provide the ideal counterpart to GW detections, offering the large-scale environmental context and statistical localisation information required to transform GW events into a powerful cosmological probe.

Both \ac{gw} sources and the \hi\ field are assumed to follow the underlying \ac{dm} field. Therefore, if \acp{bbh}, for which, on the gravitational side, we can measure the luminosity distance, are more likely to be found where there are \hi\ overdensities, we can use this information to pinpoint their redshift, which can be precisely measured with the upcoming radio surveys as SKA-Mid. This gives the luminosity distance-redshift pair for cosmological inference.

In other words, the redshift of GW detections can be constrained by building a \ac{los} redshift prior describing the probability that there is a compact binary merger, as proportional to the distribution of neutral hydrogen on that \ac{los}.
For simplicity, we assume that neutral hydrogen and merging compact binaries both trace the underlying dark matter and that their distributions are related by a scale- and redshift-independent bias.
These assumptions will need to be refined when moving to real data analysis, where additional parameters will need to be included in the model. The required refinements can in principle concern both our assumptions on the physical relation between the presence of \hi\ and GW events and also the effect that realistic systematics can have in "masking" the real \hi\ distribution. Such systematics can be the presence of residual foregrounds contamination after foreground cleaning, mis-calibration effects, non-uniform mapping of the structure and the incompleteness due to unobserved regions of the sky.

This approach is explored in both~\citep{Dupletsa:2026uqs} and \citep[][\textit{in preparation}]{Nanadoumgar-Lacroze:InPrep}, with independent underlying modelling. \cite{Dupletsa:2026uqs} make use of N-body simulations to model the underlying dark matter tracers, which provided high fidelity simulations, whilst \cite[][\textit{in preparation}]{Nanadoumgar-Lacroze:InPrep} make use of log-Normal simulations of the density fields, which have lower accuracy on small scales but can be rapidly generated, allowing the easier exploration of different physical cosmological, source and telescope models.

\begin{figure}
    \centering
    \includegraphics[width=0.85\linewidth]{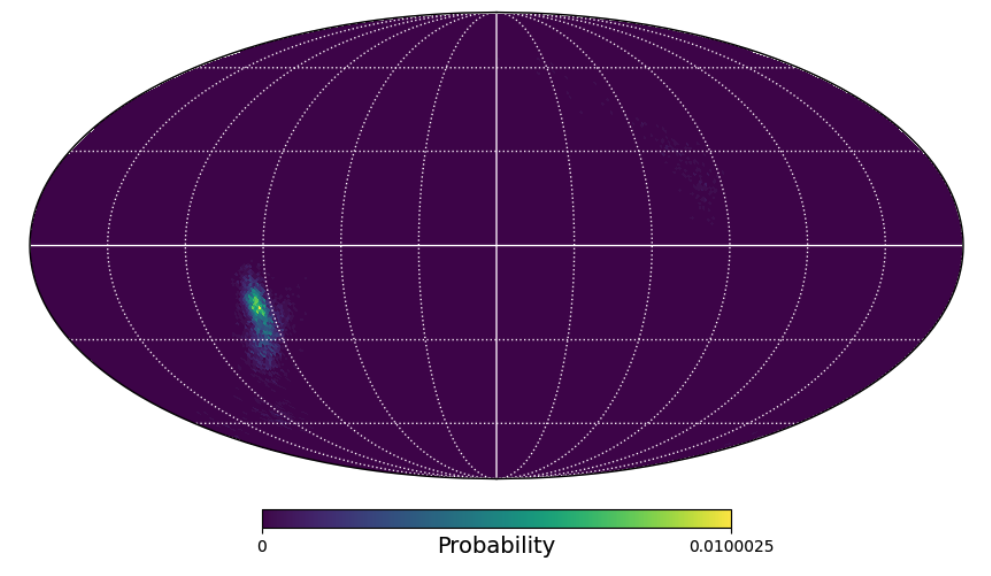}
    \caption{Pixelated sky map for a mock GW event with 90\% sky area $A_{sky}^{90\%} = 94 \space \rm deg^2$.}
    \label{fig:gw_skymap}
\end{figure}

\begin{figure}
    \centering
    \includegraphics[width=0.5\linewidth]{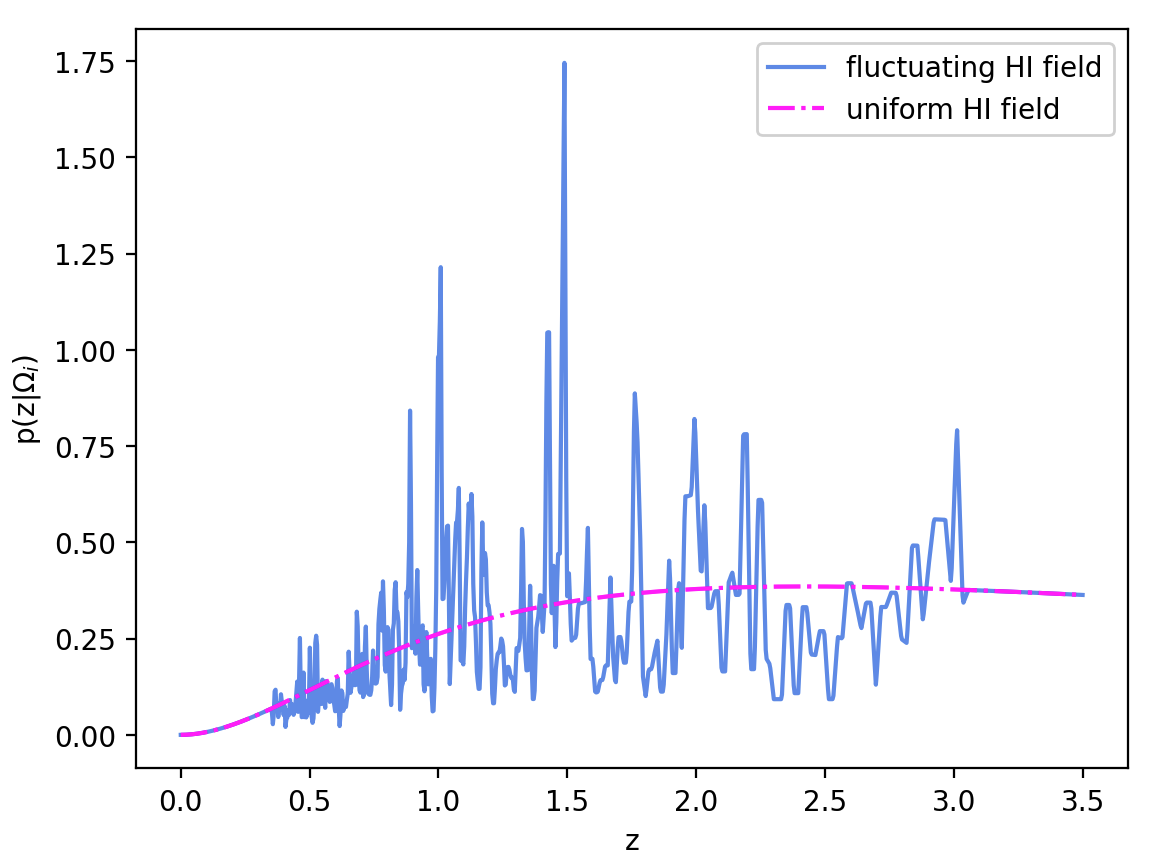}
    \caption{Line-of-sight redshift prior for a single pixel with \texttt{nside}$=128$ in the \texttt{GRIDIMP} intensity maps. We compare the redshift prior from a fluctuating \hi\ field with that of a uniform \hi\ field.}
    \label{fig:singleLoS}
\end{figure}

The sky localisation of GW events is given as pixelated sky probability maps as shown in Fig. \ref{fig:gw_skymap}.
For a two-dimensional pixel $\Omega_i$ , we construct a line-of-sight redshift prior, $P(z|\Omega_i)$, which summarizes the probability of a CBC merger for each redshift bin $z_j$, $P(z_j|\Omega_i)$.
For each pixel in the sky map, a redshift prior can be built by looking at the variation of \hi\ density along the redshift direction, defining $\mathbb{P}(z|\Omega_i)$, the CBC merger probability as a function of $z$. We have:
\begin{equation}
\begin{split}
    \mathbb{P}(z|\Omega_i)  = \sum_{j=1}^{N_{z}} f(z;z_j, \sigma_j) \mathbb{P}(z_j|\Omega_i) = \frac{1}{\mathbb{P}(\Omega_i)}\sum_{j=1}^{N_{z}}f(z;z_j, \sigma_j)\mathbb{P}(\Omega_i|z_j)\mathbb{P}(z_j), 
\end{split}
\label{eq:LoS}
\end{equation}
where:
\begin{itemize}
    \item $f(z;z_j, \sigma_j)$ is the redshift grid function, it is equal to $\delta (z-z_j)$ when the redshift coordinate $z_j$ is perfectly measured
    \item $\mathbb{P}(\Omega_i|z_j)$ is the probability that a pixel with coordinates $(\Omega_i,z_j)$ hosts a compact binary merger. It is directly computed from the \hi\ density fluctuations as given by the \hi\ intensity map.
    \item $\mathbb{P}(z_j)$ is the probability that there is a compact binary merger at redshift $z_j$ based on considerations such as the star formation rate, assuming a uniform co-moving volume and taking into account time dilation.
    \item $\mathbb{P}(\Omega_i)$ corresponds to the probability that the GW is in pixel $\Omega_i$, as described by its sky map.
\end{itemize}
Eventually, the \hi -informed redshift prior for a single GW event,  $\mathbb{P}(z)$,  is defined by summing over the redshift prior of all pixels that form part of the GW event's sky map: 
\begin{equation}
    \mathbb{P}(z) = \sum _{i=0}^{N_\Omega } \mathbb{P}(z|\Omega_i).
\end{equation}
It is possible to define those probabilities with respect to various quantities such as the \hi\ mass $\rm M_{\rm H_I}$, \hi\ density $\rm \rho_{\rm H_I}$, or \hi\ temperature $\rm T_{\rm H_I}$. Fig. \ref{fig:singleLoS} shows an example of such a redshift prior along a single line-of-sight $\Omega$.  For comparison, when using galaxies as the EM tracer, they are defined as functions of the galaxy number density and the galaxy luminosity or magnitude.

\begin{figure}[h]
    \centering
    \includegraphics[width=0.75\textwidth]{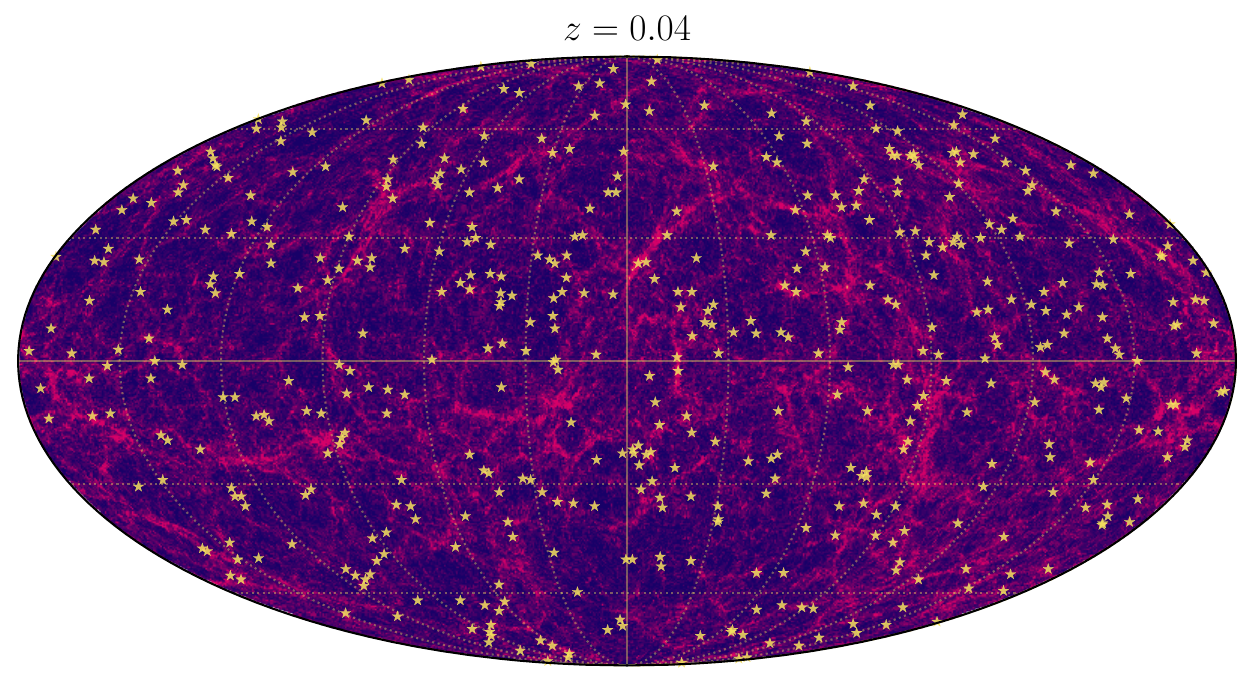}
    \caption{Illustrative example of 500 \ac{gw} events (\textit{yellow} stars) on \hi\ number density map at $z\sim 0.4$ as from the \ac{DEMNUni} N-body simulation.}
    \label{fig:gws_hi_map_uly}
\end{figure}

In addition to being fundamental to the construction of a \hi-informed redshift prior, Equation \ref{eq:LoS} is essential to the generation of a mock GW catalogue when using simulated data. For both cases here where N-body simulations and log-Normal simulations are used, the rationale is the following:
\begin{itemize}
    \item We simulate \hi\ intensity maps produced following the observational characteristics of studied \hi\ IM surveys.
    \item We create a mock GW catalogue with parameters $\{m_1, m_2, z,\rm RA, Dec.\}$. Given the dark matter distribution of our simulations, the spatial location of GW sources is drawn for each redshift shell with a probability $\mathbb{P}(\Omega_i|z_j)$ while the redshift distribution is drawn following $\mathbb{P}(z_j)$. Fig. \ref{fig:gws_hi_map_uly} shows an example of the distribution of GW events drawn from a given dark matter map, superposed to our \hi\ simulated sky.
    \item We simulate the GW parameters in detector-frame, assuming a waveform model and a GW interferometer-network. 
    \item We use the spectral siren method to infer cosmological parameters with the hierarchical Bayesian inference pipeline \texttt{icarogw} ~\citep{Mastrogiovanni:2023zbw}, adapted to allow the use of \hi\ intensity maps.
\end{itemize}

The work in \cite{Dupletsa:2026uqs} uses density maps. The starting point is the simulation of the \ac{dm} distribution at different redshift bins. Given that SKA-Mid frequency coverage will allow tomography of the $21$\,cm line of \hi\ up to $z\sim 3$, and with \ac{ET} we will be able to detect \ac{bbh} mergers much further, the analysis is set in $z\in [0, 3]$. The \ac{dm} maps come from an N-body simulation, where more realistic features, e.g., non-Gaussianities, are included.

Specifically, we use the \ac{DEMNUni} set of N-body simulations~\citep{Castorina:2015bma, Carbone:2016nzj, Parimbelli:2021mtp}. They are characterized by a Planck 2013~\citep{Planck:2013pxb} baseline flat \ac{CDM} cosmology with a box with side length of $500$\,Mpc\,$h^{-1}$ in comoving coordinates, with $2048^3$ \ac{dm} particles and no neutrino particles. A full-sky lightcone from redshift $z=0$ to redshift $z\approx8$ is then built, corresponding to a total comoving volume of approximately $3\times10^3$\,Gpc$^{3}$\,$h^{-3}$. Around $39$ billion haloes are identified through the \ac{FoF} algorithm. The \ac{dm} haloes are used as a biased tracer of the underlying continuous \ac{dm} distribution. \cite{Dupletsa:2026uqs} explores, in particular, the interplay between the \hi\ redshift prior and the uncertainties in the GW parameters measured with ET, and the role of the GW population parameters in the $H_0$ constraint.

The work in \cite[\textit{in preparation}]{Nanadoumgar-Lacroze:InPrep} uses temperature maps generated with the python package \texttt{GRIDIMP} \citep{GRIDIMP}, by drawing log-Normal realisations from the \hi\ matter power spectrum, itself derived from the \ac{dm} power spectrum. Two observational scenarios are studied: 
\begin{itemize}
    \item SKA-Mid Band 1, covering $z \in [0.3, 3]$~\citep{SKA:2018ckk};
    \item MeerKLASS UHF-Band, covering $z\in [0.4, 1.4]$ \citep{Santos_2017}.
\end{itemize}
The resulting \hi\ intensity maps are pixelated with \texttt{healpix} nside=$128$ and a frequency resolution of $1 \rm MHz$. The forecasts are made realistic by leveraging the observational and instrumental effects that are expected to most impact the quality of the maps: 1. The redshift error associated with each bin of the map; 2. The limited sky coverage of the surveys; 3. The smoothing effect of the telescope beam, which further limits the angular resolution;  4. The noise introduced by the telescope receiver that reduces the signal-to-noise ratio of the \hi\ signal; 5. The signal loss or foreground residuals, dependent on the performance of foreground cleaning technique. The differences in observational characteristics such as the diameter and number of dishes, observational time and sky coverage influence the noise level and the angular resolution achieved by the surveys.
Using the \texttt{GRIDIMP} maps, the $\rm{H_I}$ temperature $\rm T_{ H_I}(\Omega,z) = \delta \rm T_{H_I}(\Omega,z) + \bar{T}_{\rm H_I}(z)$ is used as an observable for the analysis, where $ \delta \rm T_{H_I}(\Omega,z) $ represents the temperature fluctuation at sky angle $\Omega$ and redshift $ z $ in the simulated map with respect to the mean temperature $\bar{T}_{\rm H_I}(z)$. The observational incompleteness, both in sky coverage and redshift, is mitigated by $\bar{T}_{\rm H_I}(z)$, defined in the \texttt{GRIDIMP} simulations. 

Both studies modify the CBC merger rate parametrization, essential to the cosmological inference with \texttt{icarogw}. We define the CBC merger rate, $\frac{\dd N_{\rm CBC}}{\dd t \dd z \dd \Omega}$, as the rate of  CBC mergers in detector-frame per unit redshift and unit solid angle. 
It can be written such as
\begin{equation}
\label{eq:md_rate_modified}
\frac{\dd N_{\rm CBC}}{\dd t\dd z \dd \Omega} = \mathcal{R}_0 \psi(z|\gamma,\kappa,z_p) f_{\rm H_I}(z,\Omega) \frac{1}{1+z} \frac{\dd V_c}{\dd z \dd \Omega},
\end{equation}

where $\mathcal{R}_0$ is the CBC merger rate at $z=0$, $\psi(z|\gamma,\kappa,z_p)$ is the Madau-Dickinson star formation rate \citep{Madau:2014bja}. $f_{\rm H_I}(z,\Omega)$ is the function encoding the \hi\ information proportional to the \hi\ density contrast as a function of redshift. Further details about the derivation of this expression can be found in \cite{Dupletsa:2026uqs} and \cite[\textit{in preparation}]{Nanadoumgar-Lacroze:InPrep}.

\begin{figure}[h]
    \centering
    \includegraphics[width=0.65\textwidth]{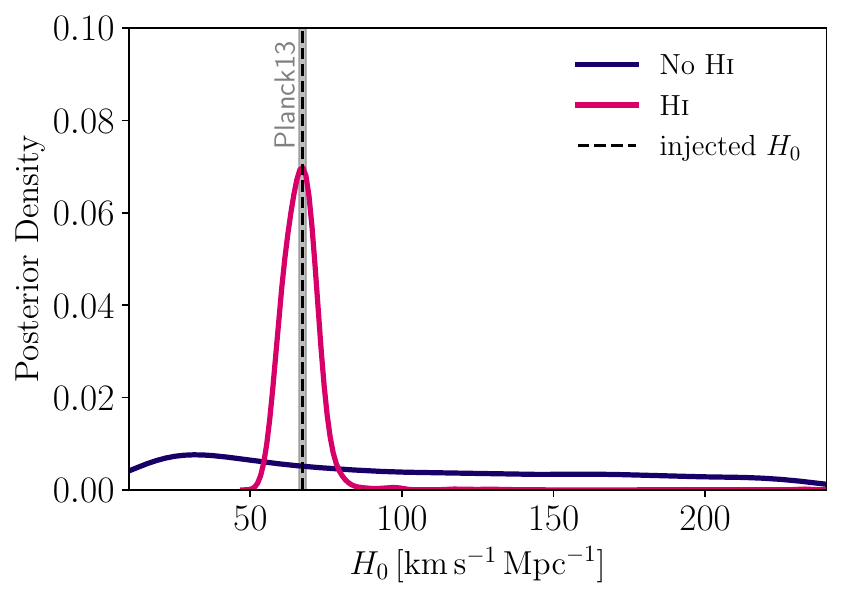}
    \caption{Reconstruction of the $H_0$ posterior using the N-body simulations and modified \texttt{icarogw} code (as described in the pipeline presented in \cite{Dupletsa:2026uqs} with $\sim 3000$ \ac{gw} events with an \ac{snr} greater than 150. We present the results using the N-body \hi\ maps from the \ac{DEMNUni} simulation to model the \ac{gw} rate, in \textit{magenta}, and results from assuming a sky-averaged distribution of \hi, in \textit{purple}.}
    \label{fig:H0_uly}
\end{figure}

In Fig.~\ref{fig:H0_uly}, we show the $H_0$ posterior obtained with $\sim 3000$ \ac{gw} events, with the \hi\ distribution as from the N-body maps using the pipeline described in \cite{Dupletsa:2026uqs}, marginalizing over all the other population-level parameters. We can see that under the simplified analysis, using the \hi\ maps, not only we manage to recover the injected $H_0$ value (the one from Planck 2013~\citep{Planck:2013pxb} used for the N-body maps simulations), but we also recover it with great precision. For comparison, we report the measurement from Planck 2013, which estimate of $H_0$ is at the percent level. 

\begin{figure}[h]
    \centering
    \includegraphics[width=0.75\linewidth]{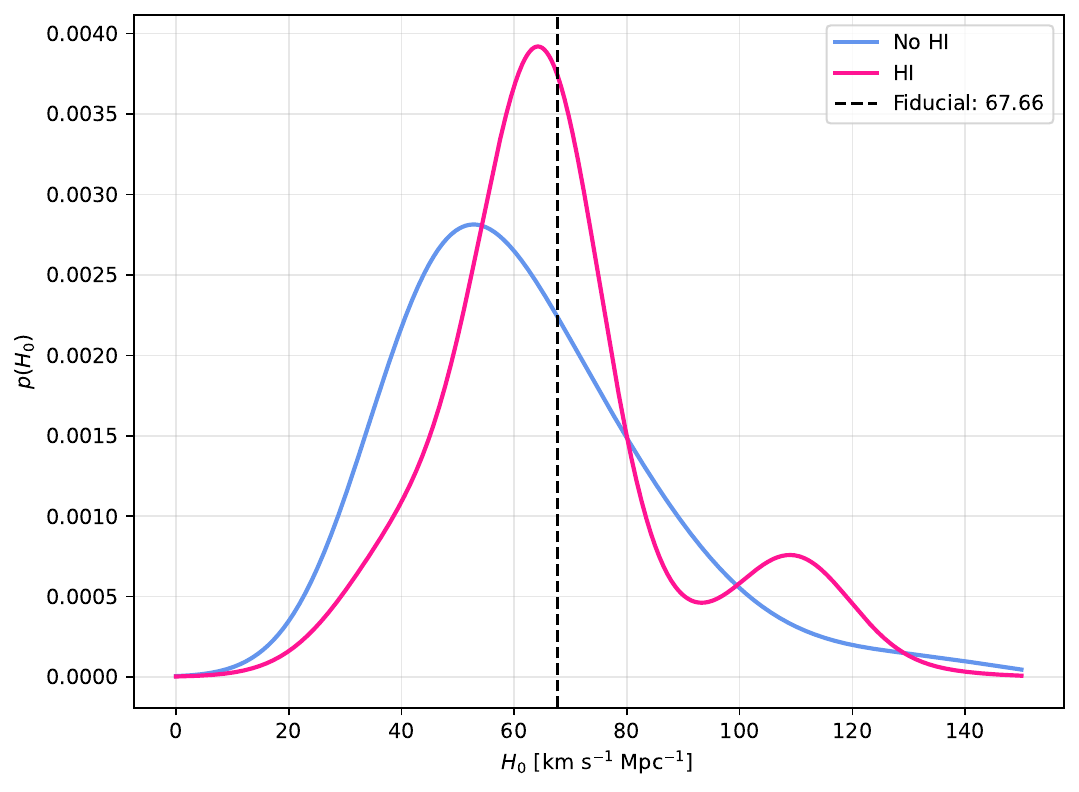}
    \caption{$\rm{H_0}$ posterior distribution marginalised over cosmological and population parameters obtained using the pipeline with log-Normal simulations presented in \cite[\textit{in preparation}]{Nanadoumgar-Lacroze:InPrep}. We show preliminary results using \texttt{icarogw} with $\sim 160$ O5-like GW events and compare the constraint on $H_0$ achieved with and without using the corresponding SKA-Mid-like $\rm{H_I}$ temperature map. For this analysis, the information from the mass spectrum, in
particular a broken power law plus a two peaks model is also considered.}
    \label{fig:H0_dounia}
\end{figure}

A similar analysis is presented in \cite[\textit{in preparation}]{Nanadoumgar-Lacroze:InPrep}, where the marginal $H_0$ posterior is obtained with about 160 GW events drawn from the LVK zero-noise O5 sensitivity curve with a \ac{snr} threshold of 12, and a SKA-Mid-like temperature map generated with log-Normal simulations from \texttt{GRIDIMP}. Cosmological and GW population parameters are inferred jointly leveraging on the interplay of the source-frame mass distribution with cosmology to avoid biases \citep{Mastrogiovanni:2021wsd} and marginalise the $H_0$ posterior from the other parameters. Please note that in this case the procedure use also the information from the mass spectrum, in particular a broken power law plus a two peaks model is considered.
In this framework, it is possible to recover the fiducial cosmological and population parameters, including $H_0$ which is measured with a 19 \% error (pink curve in Fig.~\ref{fig:H0_dounia}). This corresponds to a 10\% improvement with respect to performing an identical analysis without relying on \hi\ maps (blue curve in Fig.~\ref{fig:H0_dounia}). These results show, in particular, how observational and instrumental limitations introduced by the SKA-Mid (Band 1) and MeerKLASS (UHF-band) surveys are expected to impact the constraining power of the method. Anticipating the challenges that may arise with real data, these results show that \hi\ intensity maps could allow us to measure $H_0$ with a precision that matches and maybe exceeds the current estimate from LVK GWTC-4 \citep{LIGOScientific:2025jau}, and therefore provide a valuable data set, complementary to galaxy surveys.  The use of the information from the mass spectrum explains why even for the no \hi\ case there is constraining power. This information is instead not used in Fig.~\ref{fig:H0_uly}, justifying why there is no constraining power there despite considering more GW events.

\subsection{Other Radio Tracers as a Redshift Prior}\label{sec:radio_tracers}
\subsubsection{Continuum Galaxies}\label{subsec:continuum_gal}
SKA-Mid will also perform large area surveys of continuum radio galaxies; a sample which can provide information on the hosts of dark siren events. Though these surveys will contain large numbers of sources compared to optical surveys, and may have redshift distributions which peak at higher redshifts, they will likely not be directly usable for line of sight prior information due to their intrinsic lack of redshift information. Such redshift information will come either from cross-matching with surveys in optical and near-InfraRed bands, or other approaches such as clustering redshifts. Nonetheless, radio continuum information may provide useful information on the host probability for each galaxy through enhancing the multi-wavelength view of star formation and evolution (as discussed elsewhere in this volume e.g. \citealt{Asorey01.2026.SKA}).

\subsubsection{\hi\ Galaxies}\label{subsec:hi_gal}
Similarly, large scale SKA-Mid surveys of \hi\ line emission in galaxies could also provide a line of sight redshift prior. As discussed in \cite{Yahya:2014yva,Nasirudin01.2026.SKA} these surveys could contain $\mathcal{O}(10^6)$ galaxies with spectroscopic redshifts in the range $z < 0.5$, highly competitive with the optical spectroscopic surveys often considered for dark sirens analyses. Such surveys will also have a distinct advantage of cleaner selection functions than many optical surveys, bypassing a key systematic uncertainty in their use for dark sirens.

\section{Angular Cross-Correlations}\label{sec:cross_corr}

In addition to the map-space information used in the above sections on dark sirens, the angular cross-correlation between any pair of matter tracers, can be calculated via the angular power spectra $C_\ell$s of such tracers.
This technique is particularly powerful in this context for some key reasons.
First, given the lack of electromagnetic counterparts for BBH mergers, it allows for the calibration of the redshift distribution of GW events thanks to the high redshift precision of \hi\ IM.
Secondly, the cross-correlation of two independent cosmological probes provides intrinsic robustness to instrumental and observational systematics, as uncorrelated errors between the datasets of the different tracers naturally vanish in the cross-spectrum. 
Additionally, by taking advantage of being a multi-tracer method, cross-correlation cancels the cosmic variance from the analysis, thus improving the overall constraining power of the datasets.
Lastly, it helps to constrain the linear clustering bias of GW with respect to the underlying DM distribution, which would provide hints on the formation and evolution of the BBH population.
This formalism is well established within the cosmological community, with its first applications presented in \cite{regos_1989, Scharf_1992, Lahav_1994, Fisher_1994}.

For a given tracer \textit{X}, its matter fluctuations $\delta^X$ at a generic position in the sky $(\theta, \phi)$ and radial distance $x$ can be described as:
\begin{equation}
    \delta^X(\theta, \phi, x) = \frac{\rho^X(\theta, \phi, x) - \langle \rho^X \rangle(x)}{\langle \rho^X \rangle(x)}
\end{equation}
where $\rho^X(\theta,\phi, x)$ is the matter density and $\langle \rho^X \rangle(x)$ is the average matter density on the entire sky at radial distance $x$.
By projecting it onto the sky surface, it is possible to expand the matter fluctuations in spherical harmonics, as:
\begin{equation}
    \delta^X(\theta, \phi, x) = \sum_{\ell =0}^{+\infty} \sum_{m =-\ell}^{+\ell} a_{\ell m}^X(x)Y_{\ell m}(\theta, \phi)
\end{equation}
where $a_{\ell m}^X(x)$ and $Y_{\ell m}(\theta,\phi)$ represent, at each angular scale $\ell$ and mode $m$, the harmonic coefficients and the spherical harmonics, respectively.
The observed angular power spectrum $\tilde{C}_\ell^{XY} (x_i, x_j)$ between tracers $X$ and $Y$, at radial distances $x_i$ and $x_j$, is defined as the covariance of their angular harmonics coefficients:
\begin{equation}\label{eq:harmonics}
   \langle a_{\ell m}^X(x_i), a_{\ell' m'}^{Y^*}(x_j) \rangle = \delta_{\ell \ell'} \delta_{m m'} \tilde{C}_\ell^{XY} (x_i, x_j)
\end{equation}
with $\delta$ denoting Kronecker deltas.
Crucially, since the harmonic coefficients encode the decomposition onto spherical harmonics of the matter fluctuations, the angular power spectrum $\tilde{C}_\ell$ directly measures the covariance of these projected fluctuations on the sky.

To explicitly describe the observed angular power spectrum, we can decompose the harmonic coefficients into two separate terms, one including the signal contribution $s_{\ell m}$ and one including the noise contribution $n_{\ell m}$:
\begin{equation}
    a_{\ell m}^X(x_i) = s_{\ell m}^X(x_i) + n_{\ell m}^X(x_i)
\end{equation}
thus, by assuming that:
\begin{equation}
    \langle s_{\ell m}^X(x_i), s_{\ell' m'}^{Y^*}(x_j) \rangle = \delta_{\ell \ell'} \delta_{m m'} C_\ell^{XY} (x_i, x_j)
\end{equation}
\begin{equation}
    \langle n_{\ell m}^X(x_i), n_{\ell' m'}^{Y^*}(x_j) \rangle = \mathcal{N}_{\ell m}^{XY} (x_i, x_j)
\end{equation}
where $C_\ell^{XY} (x_i, x_j)$ is the theoretical angular power spectrum and $\mathcal{N}_{\ell m}^{XY} (x_i, x_j)$ is the noise spectrum and represents the covariance of the projected noise fluctuations on the sky; it is then possible to write the observed angular power spectrum as:
\begin{equation}
    \tilde{C}_\ell^{XY} (x_i, x_j) = C_\ell^{XY} (x_i, x_j) + \mathcal{N}_{\ell m}^{XY} (x_i, x_j)
\end{equation}
Assuming the Limber approximation, the theoretical power spectrum, describing the correlation between two tracers $(X,Y)$ at any distance $(x_i, x_j)$, is defined as:
\begin{equation}
    C_\ell^{XY} (x_i, x_j) = \frac{2}{\pi} \int_{z_{min}}^{z_{max}}  \frac{c dz}{H(z) \chi^2(z)} W^X(z, x_i) W^Y(z, x_j) \mathcal{P} \left( k = \frac{\ell+1/2}{\chi(z)}, z \right)
    \label{eq:cl_th}
\end{equation}
where $k$ is the comoving wavenumber, $\chi(z)$ is the comoving distance at redshift $z$, $H(z)$ is the Hubble parameter at $z$, $W^X(z, x_i)$ and $W^Y(z, x_j)$ are the observational window functions for each tracer at their corresponding redshift bin, $\mathcal{P}(k,z)$ is the primordial power spectrum, and $ \left[ z_{min}, z_{max} \right]$ represents the redshift range of the survey.
The window functions depend on the specific experiments of the different tracers and include contributions from the observed redshift distribution of each tracer, redshift and distance measurement errors, and tracer biases.
Note that, to track how the signal evolves with redshift, in equation~(\ref{eq:cl_th}) a tomographic approach has been applied, i.e. evaluating the distribution of each source at different redshift bins.
Crucially, equation~\eqref{eq:cl_th} provides only a schematic baseline which includes only spatial density fluctuations.
As such, the forecasts summarized below do not rely on this approximation, as a fully relativistic, non-Limber approach, computed via \texttt{Multi\_CLASS} (\cite{Bellomo20:multiclass}), is adopted.

Given that the noise is intrinsically linked to the observational setup for each source, the noise spectrum has also to be discussed separately for each tracer.
In the specific case of discrete sources, such as GW signals, the noise spectrum can generally be defined as:

\begin{equation}
    \mathcal{N}_\ell^{XY} (x_i, x_j) = \frac{4 \pi f^X_{\mathrm{{sky}}}}{N^X(x_i)} \frac{\delta_{ij}}{\left[ \mathcal{W}^X_\ell(x_i) \right]^2 }
    \label{eq:noise}
\end{equation}
where $f^X_{\mathrm{sky}}$ is the fraction of the sky covered by the observation of tracer $X$, $N^X(x_i)$ represents the number of events of tracer $X$ in the $i$--th redshift bin, $\delta_{ij}$ are Kronecker deltas, and $\mathcal{W}^X_\ell(x_i)$ is the beam window function, which accounts for the localization errors due to the instruments for a given multipole $\ell$ at the $i$--th redshift bin.
In contrast, hydrogen IM measurements do not resolve each source individually, but instead observe the overall emission, thus not allowing for the application of equation~(\ref{eq:noise}).
Due to the nature of these unresolved observations, strictly linked to the observational setup, the noise spectrum must be discussed on a tracer-by-tracer basis. 

\begin{figure}[t!]
    \centering
    \includegraphics[width=1.0\textwidth]{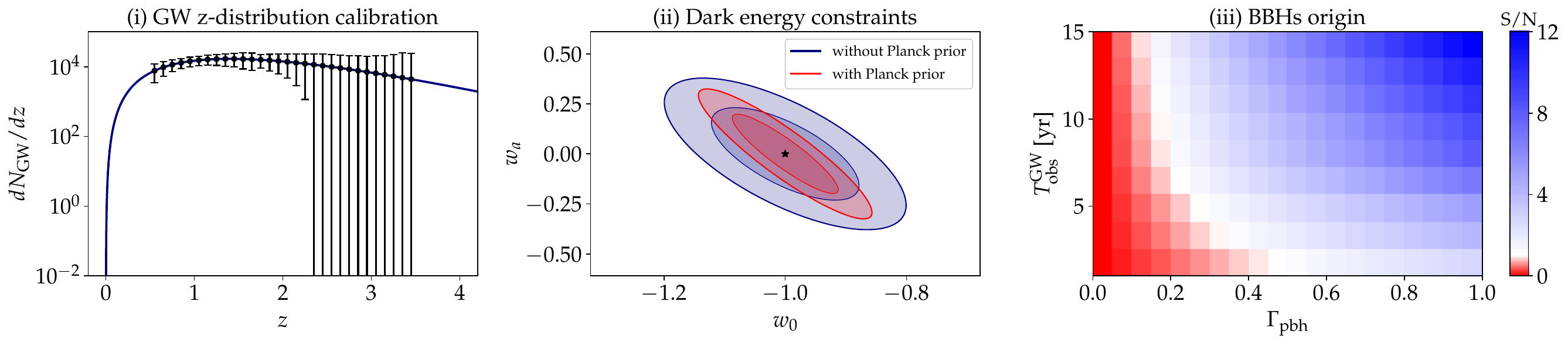}
    \caption{Summary results from \cite{Scelfo:2021fqe}, assuming $f_{\rm sky}=0.5$. {\it (i) Left panel}: constraints (absolute error bars) on the statistical z-distribution $dN_{\rm GW}/dz$ of resolved BBHs merger events, without any assumption on the underlying cosmology or astrophysical models. Assuming $T^{\rm GW}_{\rm obs} = 15$yr. {\it (ii) Center panel}: $1\sigma - 2 \sigma$ Fisher contours on dark energy parameters $\{w_0, w_a\}$, either with and without a Planck prior on the other cosmological parameters considered in the analysis. {\it (iii) Right panel}: S/N ratios for the detection of a fraction $\Gamma_{\rm pbh}$ of BBH mergers of primordial origin (assuming binary formation of PBHs binaries at late time, and astrophysical BBHs as ground truth fiducial model). Different observation times $T^{\rm GW}_{\rm obs}$ for ET are considered. See main reference for a full treatment of different scenarios.}
    \label{fig:scelfo22}
\end{figure}

The authors of \citep{Scelfo:2021fqe} investigate the cross-correlation signal between resolved GW BBHs sources and \hi\ intensity mapping, focusing on the combined capabilities of ET \citep{Sathyaprakash:ET,Maggiore:2019uih,Branchesi:2023mws,ET:2025xjr} and SKA-Mid respectively, tackling three main topics (listed below).
The study adopts a fully relativistic tomographic approach based on the angular power spectra $C_{\ell}^{\rm GW \times \hi}(z_i,z_j)$ of equation \eqref{eq:cl_th}, including density, redshift-space distortions, lensing magnification and potential (GR) terms.
The main noise contributions included in the analysis are the shot noise for the GW sources, and the single-dish thermal instrumental noise together with residual foregrounds for the \hi\ IM.
The angular spectra are computed using \texttt{Multi\_CLASS} (\cite{Bellomo20:multiclass}) and forecasts are obtained through a Fisher matrix approach covering the redshift range $0.5 < z < 3.5$ with a binning of $\Delta z_{\rm GW}=1.0$ and $\Delta z_{\rm IM}=0.1$, assuming ET in its $\Delta$ configuration and different observational campaigns from 1\,year up to 10\,years.
In the following, details on the three applications explored in \cite{Scelfo:2021fqe} are provided:
\begin{enumerate}[i)]
    \item \textit{Statistical redshift calibration of GW events}. In the absence of electromagnetic counterparts, GW luminosity distances yield poorly constrained redshifts. The ${\rm GW\times \hi}$ cross-correlation provides a tomographic mapping between the poorly localized GWs and the precisely localized \hi\ IM field, allowing to infer the overall statistical redshift distribution $dN_{\rm GW}/dz$. Modeling $dN_{\rm GW}/dz$ as a piecewise function with a different amplitude $A_i$ at each \hi\ bin, Fisher forecasts show relative errors  $\sigma_{A_i} / A_i < 1$ up to $z \approx 2$ for $f_{\rm sky} \geq 0.5$ and multi-year ET operativity $T^{\rm GW}_{\rm obs}$, enabling GW sources calibration without any assumption on cosmology or astrophysics (see e.g. left panel of Fig.~\ref{fig:scelfo22}).
    \item \textit{Constraints on dynamical dark energy}. As an example of cosmological application, extending the parameter space to a total of 10 (cosmological, bias and IM foreground parameters), the authors forecast errors on the $(w_0, w_a)$ parameters describing the dark-energy equation of state (see central panel of figure \ref{fig:scelfo22}) to be comparable with BOSS-Planck constraints. Cross-correlations help decorrelating degeneracies and overcome systematics, also providing a fundamental validation of IM-only results, which could be affected by unknown systematic errors.
    \item \textit{Probing the nature of BBH progenitors}. The clustering bias of resolved GW sources encodes crucial information on BHs origins. Astrophysical BBH mergers are linked to stellar populations in massive star-forming and high-bias halos, highly cross-correlating with \hi\ too. Primordial black holes (PBH) are theorized to be formed in the early universe (\cite{hawking:pbh, carr:pbh}) and are expected to trace the dark matter field with a lower bias (\cite{bird:pbhasdarkmatter, Scelfo18:gwxlss, Bosi23:gwxlss}) whose value depends on the time of binary capture. Forecasts show that $\mathrm{GW \times \hi}$ measurements could detect a PBH merger fraction ($\Gamma_{\rm PBH}$) down to $\sim 10\%$, corresponding to a measurable deviation in the cross-bias amplitude of $C_\ell^{\rm GW, HI}$ (see right panel of figure \ref{fig:scelfo22}).
\end{enumerate}
Finally, the authors of \cite{scelfo23:mg} have shown that with a similar approach, focusing mainly on lensing and angular clustering, the $\mathrm{ET \times SKAO}$ cross-correlation can shed light on beyond-GR signatures and Modified Gravity.

An extension to the Fisher-matrix approach presented in \cite{Scelfo:2021fqe} and \cite{scelfo23:mg} is explored in \cite[\textit{in preparation}]{SchulzInPrep}, where a likelihood-based Bayesian analysis has been carried out.
This extension allows for a more robust cosmological inference from the combined observations of \ac{ET} and \ac{skao}.

In the previous works listed here it is assumed that either the redshift of the GW is known via the host galaxy or a photometric redshift-like distribution can be built from the potential host galaxies in the field where the event came from. The authors in \citet{2023JCAP...08..050F} take a different approach and consider the only distance that can be inferred from GW events, the luminosity distance. They then compute the full number counts expression in luminosity-distance space and adapt \texttt{CAMB}\footnote{https://github.com/cmbant/CAMB} to allow for luminosity-distance space tracers of dark matter. They also show that the $C_\ell$ can be up to 30\% different from the redshift space expression. The same authors recently looked at the angular power spectrum of correlations between radio tracers and GW events \citep{Zazzera:2025ord}. Their focus was on synergies with future third-generation GW detectors like the Einstein Telescope (ET) and Cosmic Explorer (CE) which are expected to detect millions of BBH mergers. For the SKAO they considered \hi\ intensity mapping, an \hi\ galaxy surveys and a radio continuum galaxy catalogue. The authors take a full multi-tracer approach and conclude that such synergies can greatly inform about GW  astrophysical parameters, namely their clustering bias. Of all, synergies with the \hi\ IM band 1 wide area survey is the one which can provide the tightest constraints on GW parameters as well as on standard cosmological parameters. Additionally, they explored a triple cross-correlation using GWs, IM, and photometric galaxies. The three
multi-tracer method enhances constraints on the magnification lensing effect, achieving percent-level precision, and allows for a
measurement of the Doppler effect with approximately 15\% uncertainty. Such uncertainty can only be attained when adding the radio information and will give a new stage to perform null tests of
GR in cosmological scales.

\section{Conclusions}\label{sec:conclusions}
In this chapter, we highlighted the synergy between radio surveys and GW observations in the context of cosmological analyses.
With the increase in the number of detections of BBHs, BNSs, and NSBHs expected from third-generation GW observatories, the role of radio observatories will become more and more central. In particular, the SKAO will provide valuable support for a wide range of redshifts, going potentially beyond the reach of the optical galaxy surveys. 

The identification of a variety of EM counterparts will provide the redshift information necessary to use bright sirens efficiently for cosmology. In an idealised scenario, we found that, with a few tens of events, it will be possible to constrain $H_0$ and $\Omega_m$ with comparable precision of current SNe and BAO surveys and even to explore evolving dark energy scenarios. 

In the context of dark sirens, neutral hydrogen intensity mapping  (\hi\ IM) observations of the large scale structure will provide a unique opportunity to improve the priors of hierarchical Bayesian inference and to explore angular cross-correlation analyses. \hi\ IM, with its relative low resolution but high survey speed and very precise redshift information, is already mapping large volume of the universe and will provide, thanks to the SKAO, competitive and complementary measurements to optical galaxies.
Combining it with current and next-generation GW detector networks using the spectral siren method, will allow us to measure cosmological parameters such as the Hubble constant $H_0$ with great accuracy, even including uncertainties arising from the detection of GWs and SKAO instrumental limitations. 

For future surveys, using cross-correlations will also be crucial to constrain cosmological parameters: cross-correlation methods have the potential to probe the nature of BBH progenitors and to shed light on dynamical dark energy.

\section*{Author List Ordering}
The authors of this chapter have contributed equally to the conception and the writing, and are listed alphabetically.

\section*{Acknowledgements}
JF thanks the support of FCT - Fundação para a Ciência e a Tecnologia through national funds by these grants: UID/04434/2025 (DOI 10.54499/UID/04434/2025) and 2023.15069.PEX. JF acknowledges the support from FCT in the form of work through the Scientific Employment Incentive program (reference 2020.02633.CEECIND/CP1631/CT0002)
M.~S. is supported by the French government through the France 2030 investment plan managed by the National Research Agency (ANR), as part of the Initiative of Excellence Université Côte d’Azur under reference number ANR- 15-IDEX-01 and by the French Programme National de Cosmologie et Galaxies (PNCG project CIMES).
A.~C. and S.~M are supported by ERC grant GravitySirens  101163912. Funded by the European Union.
SC acknowledges support from the UKRI Stephen Hawking Fellowship (grant reference EP/U536751/1). 
GS is partially supported by INFN INDARK grant. GS is partially supported by Next Generation EU programme (PNRR).
This work is partially supported by the Spanish MCIN/AEI/10.13039/501100011033 under the Grants No. PID2020-113701GB-I00, PID2023-146517NB-I00 and CEX2024-001441-S, some of which include ERDF funds from the European Union, and by the MICINN with funding from the European Union NextGenerationEU (PRTR-C17.I1) and by the Generalitat de Catalunya. IFAE is partially funded by the CERCA program of the Generalitat de Catalunya.
This project has received funding from the European Union’s Horizon Europe research and innovation programme under the Marie Skłodowska-Curie grant agreement No.10181337.

\bibliographystyle{abbrvnat-maxbibnames4}
\bibliography{chapter} 

\end{document}